\documentstyle[aps,prb]{revtex}
\begin{document}
\title{Impact of criticality and phase separation on the spin dynamics \\
of the one-dimensional $t$-$J$ model}

\author{Shu Zhang and Gerhard M\"uller}
\address{Department of Physics, The University of Rhode Island, Kingston,
Rhode Island 02881-0817}
\author{Joachim Stolze}
\address{Physikalisches Institut, Universit{\"a}t Bayreuth, 95440 Bayreuth,
Germany\\ and Institut f\"ur Physik, Universit\"at Dortmund,
44221 Dortmund, Germany}

\date{\today}
\maketitle
\begin{abstract}
The recursion method is used to determine the $T=0$ spin dynamic structure
factor $S_{zz}(q,\omega)$ in the Luttinger liquid state and in the
phase-separated state of the one-dimensional $t$-$J$ model.
As the exchange coupling increases from zero, the dispersions and
line shapes of the dominant spin excitations are observed to undergo
a major metamorphosis between the free-fermion limit and the onset of
phase separation.
The familiar two-spinon spectrum of the Heisenberg antiferromagnetic chain
emerges gradually in the strongly phase-separated state.
\end{abstract}

\twocolumn

The $t$-$J$ model for strongly correlated electrons has been widely used
as a prototypical system for the study of the interplay between charge and
spin degrees of freedom.\cite{tj}
For a one-dimensional (1D) lattice of $N$ sites the Hamiltonian reads
\begin{eqnarray}
H_{t\text{-}J}=&&-t\sum_{\sigma =\uparrow ,\downarrow }\sum_{l=1}^N\left\{
\tilde c_{l,\sigma }^{\dagger }\tilde c_{l+1,\sigma }\ +\ \tilde c%
_{l+1,\sigma }^{\dagger }\tilde c_{l,\sigma }\right\} \nonumber \\
&&+\
J\sum_{l=1}^N\left\{ {\bf S}_l\cdot {\bf S}_{l+1}-\frac 14n_ln_{l+1}\right\}
\label{1}
\end{eqnarray}
with $\tilde c_{l,\sigma } = c_{l,\sigma }(1-n_{l,-\sigma })$, $n_l
= n_{l,\uparrow }+n_{l,\downarrow }$, $n_{l,\sigma } =
c_{l,\sigma }^{\dagger }c_{l,\sigma }$, $S_l^z = \frac 12(n_{l,\uparrow
}-n_{l,\downarrow })$, and
$S_l^{+} = \tilde c_{l,\uparrow }^{\dagger }\tilde c_{l,\downarrow }$.
Here we consider the quarter-filled-band case ($N_e = N/2$ electrons).

The hopping term amounts to an effectively repulsive force between
electrons on nearest-neighbor sites,\cite{hop} and the exchange term
represents an attractive force if these two electrons have opposite spins.
As the exchange interaction increases from zero,
the ground state of $H_{t\text{-}J}$ undergoes a transition,
at $J/t \simeq 3.2$, from a Luttinger liquid state to a phase-separated
state.\cite{OLSA91,AW91,HM91}
In the Luttinger liquid state, both the charge correlations (at $q=\pi$)
and the spin correlations (at $q=\pi/2$) are critical.
The transition, which is driven by the spin coupling,
produces charge long-range order (at $q=0$)
combined with a new type of spin criticality (at $q=\pi$).

The focus of this study is on the frequency-dependent spin fluctuations
of $H_{t\text{-}J}$ as they manifest themselves in the $T=0$ spin dynamic
structure factor
\begin{equation}
S_{zz}(q,\omega )=\frac 1N \sum_{l,m}e^{-iqm}\int_{-\infty
}^{+\infty }dt e^{i\omega t}\langle S_l^z(t)S_{l+m}^z\rangle.
\label{2}
\end{equation}
For the calculation of this quantity, we employ the recursion
method\cite{rm} in combination with a {\em strong-coupling
continued-fraction analysis}. \cite{VM94,VZSM94}
The recursion method in the present context is based on an orthogonal
expansion of the wave function $|\Psi_q^z(t)\rangle$ = $S_q^z(-t)|G\rangle$,
where $S_q^z$ = $N^{-1/2}\sum_l e^{iql}S_l^z$ is the spin fluctuation
operator, and $|G\rangle$ is the finite-size ground-state wave function of
(\ref{1}).
After some intermediate steps, the algorithm produces a sequence of
continued-fraction coefficients
$\Delta^{z}_1(q), \Delta^{z}_2(q), \ldots $
for the relaxation function,
\begin{equation}
c_0^{zz}(q,z) = \frac{1}{\displaystyle z + \frac{\Delta^{z}_1(q)}
{\displaystyle
z +  \frac{\Delta^{z}_2(q)}{\displaystyle z + \ldots }  }  } \;\;,
\label{3}
\end{equation}
which is the Laplace transform of the symmetrized correlation function
$\Re\langle S_q^z(t)S_{-q}^z \rangle/\langle S_q^zS_{-q}^z\rangle$.
The $T=0$ spin dynamic structure factor (\ref{2}) is then obtained from
(\ref{3}) via the relation
\begin{equation}
S_{zz}(q,\omega) =  4\langle S_q^zS_{-q}^z\rangle\Theta(\omega) \lim
 \limits_{\varepsilon
\rightarrow 0}  \Re [c_{0}^{zz} (q, \varepsilon - i\omega)] .
\label{4}
\end{equation}

In a previous paper\cite{VZMS95} we have used the recursion method together
with a {\em weak-coupling continued-fraction analysis} for a study of the
charge dynamic structure factor in the Luttinger liquid phase of the $t$-$J$
model.
The spin dynamics poses a far greater challenge.
Very few explicit results seem to exist.\cite{sd}
One key spin dynamical property in the Luttinger liquid phase can be
inferred from the asymptotic behavior of the static spin correlations
as proposed in previous work:\cite{AW91,HM91}
\begin{equation}
\langle S_l^zS_{l+m}^z\rangle \sim
B_1\frac 1{m^2} + B_2\frac{\cos(\pi m/2)}{m^{\eta_\rho/4+1}}\;\;,
\label{5}
\end{equation}
where $\eta_\rho$ is the exponent which also governs the algebraic decay
of the ($q=\pi$)-oscillations, $\sim$ $\cos(\pi m)/m^{\eta_\rho}$,
in the static charge correlation function $\langle n_l n_{l+m}\rangle$.
This exponent is known to assume the value $\eta_\rho=2$ in the free-fermion
limit $(J/t=0^+)$.
It increases linearly, $\Delta\eta_\rho \simeq 0.40J/t$, in the
weak-coupling regime,\cite{OLSA91,VZMS95}
assumes the value $\eta_\rho \simeq 3.4$ at $J/t=2$ (supersymmetric
case),\cite{KY90} and diverges at the endpoint, $J/t \simeq 3.2$,
of the Luttinger liquid state.\cite{OLSA91}

The oscillatory term in (\ref{5}) implies that the dynamically relevant
excitation spectrum of $S_{zz}(q,\omega)$ is gapless at $q=\pi/2$.
The spectral-weight distribution of the spin dynamic structure factor at
this critical wave number can then be predicted (under mild assumptions)
to have a singularity of the form
\begin{equation}
S_{zz}(\pi/2,\omega) \sim \omega^{\eta_\rho/4-1} \;.
\label{6}
\end{equation}
The infrared exponent starts out negative in the free-fermion limit,
$\sim \omega^{-1/2}$, increases monotonically with increasing $J/t$,
and then diverges at $J/t\simeq 3.2$, where
phase separation sets in.
A landmark change in $S_{zz}(\pi,\omega)$ is expected to occur at the point
where the infrared exponent switches sign (from negative to positive).
This happens for $\eta_\rho =4$, which corresponds to the coupling
strength $J/t\simeq 2.3$.

Our results for $S_{zz}(q,\omega)$ indicate that the Luttinger liquid
phase of the $t$-$J$ model can be divided into two regimes with distinct
spin dynamical properties.
As a representative result of the first regime $(0 < J/t \lesssim 1)$,
we show in Fig. 1 $S_{zz}(q,\omega)$ as a continuous function of $\omega$
and a discrete function of the $q$-values realized in a system of
$N=12$ sites with coupling strength $J/t=0.1$.

All results presented here have been calculated via a strong-coupling
continued-fraction reconstruction from the coefficients
$\Delta_1, \ldots, \Delta_6$ and a Gaussian terminator.
The $\Delta_k$'s have been extracted via the recursion method
from the ground-state wave function for $N=12$, which in turn has been
computed via the conjugate-gradient method.
The entire procedure was explained in Ref. \onlinecite{VM94}.

Throughout the Brillouin zone except at small $q$ we observe a well-defined
dynamically relevant spin mode with a $|\cos q|$-like dispersion as
indicated by the full circles.
Near the critical wave number, $q= \pi/2$, the function
$S_{zz}(q,\omega)$ may exhibit a power-law divergence of the form
$\sim [\omega^2 - c^2(\pi/2-q)^2]^{(\eta_\rho/4-1)/2}$,
similar to what has been observed at the critical wave number in other
Luttinger liquids.\cite{lutt}
At long wavelengths the spectral weight in $S_{zz}(q,\omega)$
is concentrated at fairly low frequencies.
Data for longer chains will be needed for a quantitative analysis of
the spin dispersions at small $q$ in this regime.

As the exchange coupling increases toward $J/t \simeq 1$, the following
changes in the spectral-weight distribution of $S_{zz}(q,\omega)$ can be
identified:

(i) The amplitude of the $|\cos q|$-like dispersion grows with increasing
$J/t$.
The gradual upward shift of the peak position in $S_{zz}(q,\omega )$
is accompanied by a significant increase in linewidth.
For the $q=\pi$ spin mode this is contrary to what one expects under the
influence of an antiferromagnetic exchange interaction of
increasing strength.
That trend changes at stronger coupling as we shall see.

(ii) The intensity of the central peak in $S_{zz}(\pi/2,\omega)$ weakens
gradually.
This is illustrated in the main diagram of Fig. 2.
The peak turns shallow and then vanishes altogether.
This property of our data reflects the gradual weakening and ultimate
disappearance of the power-law divergence (\ref{6}), given the
(approximately known) $J/t$-dependence of the infrared exponent.

As the coupling strength increases past the value $J/t\simeq 1$,
the spin modes which dominate $S_{zz}(q,\omega)$ in the first
regime of the Luttinger liquid phase broaden rapidly and lose their
distinctiveness.
There is a crossover region between the first and second
regimes, which roughly comprises the coupling range
$1\lesssim J/t\lesssim 2$.
Over that range, the spectral weight in $S_{zz}(q,\omega)$ is distributed
over a broad structure with rapidly shifting peaks.

At the end of the crossover region, a new type of spin mode with an
entirely different kind of dispersion has gained prominence in
$S_{zz}(q,\omega)$, and it stays dominant throughout the
remainder of the Luttinger liquid phase.
This is illustrated in Fig. 3 by a plot of $S_{zz}(q,\omega)$ for
a $J/t$-value near the onset of phase separation.
The representation is the same as in Fig. 1 except for the different
frequency scale.

The dispersion of these new spin modes,  as shown by the full circles,
does no longer have a soft mode at $q=\pi/2$.
At $J/t \sim 2.0$ the dispersion has a $|\sin(q/2)|$-like shape, i.e.
a smooth maximum at $q=\pi $ and a tendency to approach zero linearly as
$q\rightarrow 0$.
As $J/t$ increases toward the transition point, the peak positions in
$S_{zz}(q,\omega)$ gradually shift to lower values of $\omega/J$
(with $t$ held fixed).
The shift proceeds more rapidly at $q$ near $\pi$ than at smaller $q$,
which has the consequence that the maximum in the dynamically relevant spin
dispersion starts to move away from $q=\pi$ toward $q=\pi/2$.
This trend  from a $|\sin(q/2)|$-like toward a $|\sin q|$-like dispersion
persists in the phase-separated state $(J/t > 3.2)$.

The evolution of the spectral-weight distribution in $S_{zz}(\pi,\omega)$
from a well-defined mode at $\omega/J \simeq 1.4$ near the onset of
phase separation toward a soft mode in the strongly phase-separated
state is shown in the inset to Fig. 2.
By contrast, the spectral-weight in $S_{zz}(\pi/2,\omega)$, which has been
slowly shifting toward lower values of $\omega/J$ in the second
regime of the Luttinger liquid phase, now starts to move back out to
higher frequencies in the phase-separated state (not shown here).
The single-peak structure grows taller, the linewidth shrinks somewhat,
and the peak position settles at $\omega/J \simeq 1.57$.

All these properties and trends reflect or foreshadow the much better
understood $T=0$ dynamical properties of the 1D $s=1/2$ Heisenberg
antiferromagnet -- the end product of the $t$-$J$ model in the
completely phase-separated state.
The spin dynamic structure factor $S_{zz}(q,\omega)$ under these
circumstances is known to be dominated by a continuum of two-spinon
excitations with a lower boundary $\epsilon_L(q)$ = $(\pi/2)J|\sin q|$,
where the spectral-weight distribution has a divergent singularity,
and an upper boundary $\epsilon_U(q)$ = $\pi J|\sin(q/2)|$, where the
spectral-weight distribution becomes very small.\cite{VM94,VZSM94,MTBB81}

This work was supported by the U.S. National Science
Foundation, Grant DMR-93-12252. Computations were carried out on
supercomputers at the National
Center for Supercomputing Applications, University of Illinois at
Urbana-Champaign.

\pagebreak

\begin{figure}
\caption[one]
{Spin dynamic structure factor $S_{zz}(q,\omega)$ at $T=0$ for
$q=2\pi l/12, l=0,\ldots,6$ of the 1D $t$-$J$ model with $t=1$ and $J=0.1$
in the Luttinger liquid phase near the free-fermion limit.}
\label{F1}
\end{figure}

\begin{figure}
\caption[two]
{Line shape at $q=\pi/2$ of the spin dynamic structure factor
$S_{zz}(q,\omega)$ at $T=0$ for various values of $J$ in the Luttinger
liquid state of the 1D $t$-$J$ model with $t=1$ (main plot).
Line shape at $q=\pi$ for various values of $J$ in the
phase-separated state (inset).}
\label{F2}
\end{figure}

\begin{figure}
\caption[three]
{Spin dynamic structure factor $S_{zz}(q,\omega)$ at $T=0$ for
$q=2\pi l/12, l=0,\ldots,6$ of the 1D $t$-$J$ model with $t=1$ and $J=2.75$
in the Luttinger liquid phase near the onset of phase separation.}
\label{F3}
\end{figure}


\begin{references}
\bibitem{hop} At $J/t=0$ it is more likely that the site adjacent
to an electron is occupied by a hole than by another electron.
\end{references}
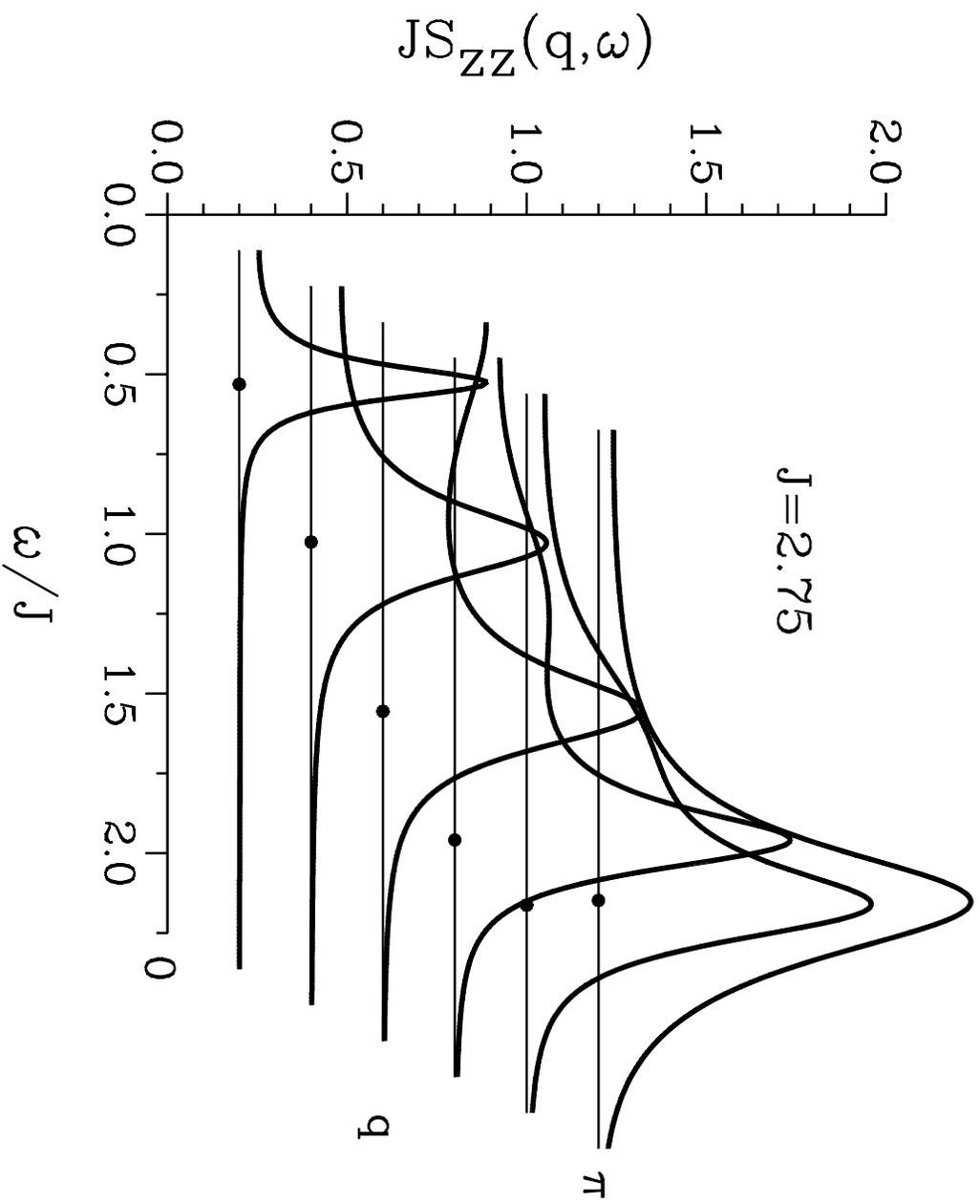
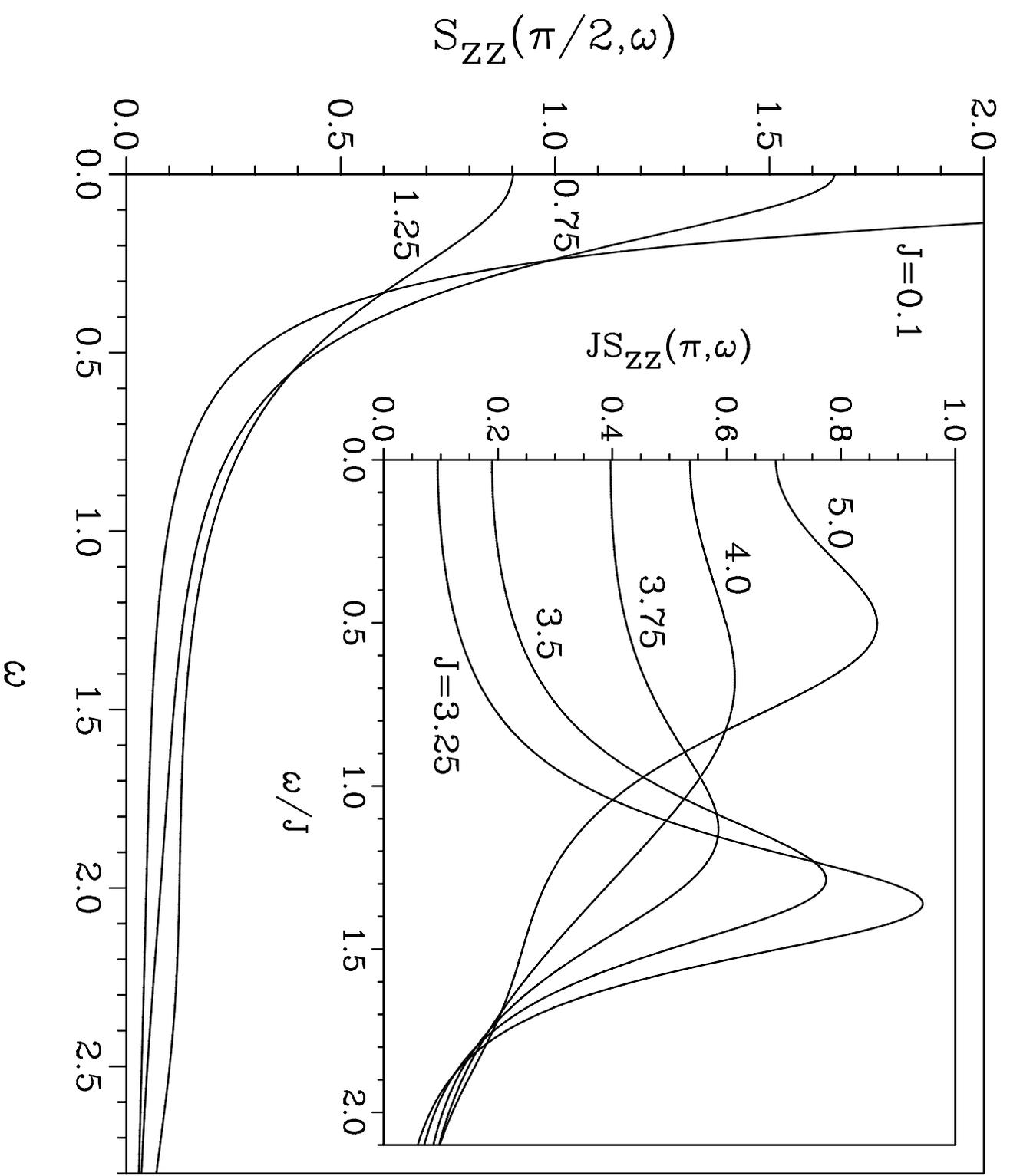
\end{document}